\documentclass[letterpaper]{article}
\usepackage{graphicx}
\usepackage{dcolumn}
\usepackage{amsmath}
\usepackage{epsfig}
\long\def\inst#1{\par\nobreak\kern 4pt\nobreak
    {\itshape #1}\par\vskip 10pt plus 3pt minus 3pt}
\RequirePackage{xspace}
\usepackage{relsize}

\def\babar{\mbox{\slshape B\kern-0.1em{\smaller A}\kern-0.1em
    B\kern-0.1em{\smaller A\kern-0.2em R}}}
\def\Kbar    {\kern 0.18em\overline{\kern -0.18em K}{}\xspace}

\def\Kz      {\ensuremath{K^0}\xspace}
\def\Kzb     {\ensuremath{\Kbar^0}\xspace}
\def\KzKzb   {\ensuremath{\Kz {\kern -0.16em \Kzb}}\xspace}

\def\Ks     {\ensuremath{K_S}\xspace}
\def\Kl     {\ensuremath{K_L}\xspace}
\def\KsKs   {\ensuremath{\Ks {\kern -0.16em \Ks}}\xspace}
\def\KlKl   {\ensuremath{\Kl {\kern -0.16em \Kl}}\xspace}
\def\KsKl   {\ensuremath{\Ks {\kern -0.16em \Kl}}\xspace}
\def\KlKs   {\ensuremath{\Kl {\kern -0.16em \Ks}}\xspace}
\def\Dbar    {\kern 0.18em\overline{\kern -0.18em D}{}\xspace}
\def\Bbar    {\kern 0.18em\overline{\kern -0.18em B}{}\xspace}

\def\Bz      {\ensuremath{B^0}\xspace}
\def\Bzb     {\ensuremath{\Bbar^0}\xspace}
\def\BzBzb   {\ensuremath{\Bz {\kern -0.16em \Bzb}}\xspace}
\def\Bu      {\ensuremath{B^+}\xspace}
\def\Bub     {\ensuremath{B^-}\xspace}

\def\BpBm    {\ensuremath{\Bu {\kern -0.16em \Bub}}\xspace}

\def\pbar    {\kern 0.18em\overline{\kern -0.18em p}{}\xspace}
\def\pp      {\ensuremath{p\pbar}\xspace}
\def\np      {\ensuremath{n\pbar}\xspace}
\def\lbar    {\kern 0.18em\overline{\kern -0.18em \Lambda}{}\xspace}
\def\pl      {\ensuremath{p\lbar}\xspace}

\newcommand{\optbar}[1]{\shortstack{{\tiny (\rule[.4ex]{1em}{.1mm})}
  \\ [-.7ex] $#1$}}
\def\BorBbar    {\kern 0.18em\optbar{\kern -0.18em B}{}\xspace}
\def\DorDbar    {\kern 0.18em\optbar{\kern -0.18em D}{}\xspace}
\def\KorKbar    {\kern 0.18em\optbar{\kern -0.18em K}{}\xspace}

\def\pep2{PEP-II}
\mathchardef\Upsilon="7107
\def\Y#1S{\ensuremath{\Upsilon{(#1S)}}\xspace}

\begin{document}

\title{ \large \bfseries \boldmath
Proposal for Direct Search for Strongly Bound States of $\pp$,
$\np$ Systems with High Intensity and Collective $\pbar$ beam }

\author{Hai-Bo Li, Mao-Zhi Yang, and Chang-Zheng Yuan \\
 CCAST (World Laboratory), P.O.Box 8730, Beijing 100080, China  \\
 Institute of High Energy Physics, P.O.Box 918, Beijing  100049, China }


\date{\today}
\maketitle

\begin{abstract}
In this letter, we discuss the possibility to look for the direct
evidence of the existence of $\pp$ and $\np$ bound states.
Measurement of the single $\gamma$ ray from $\pp$ and $\np$
systems at rest can directly confirm whether $X(1860)$ and
$X(1835)$ are the resonances which are strongly coupled to $\pp$.
In addition to the neutral candidate, a charged resonance $X^-$ is
also proposed to be searched for in $\np$ channel. We find that
the data from the Crystal Barrel experiment at LEAR/CERN can be
used to confirm the $X(1835)$ observed by BES Collaboration.  The
possibility of measuring the $\gamma$ spectrum below 100 MeV at
the new experiment with cold high intensity $\pbar$ beam at GSI is
discussed. These new techniques can be used to probe the structure
of the $X(1860)$ and $X(1835)$ in the future.
\end{abstract}

\vspace{0.5cm}
\vspace{1.5cm}




In the last few years, BES Collaboration observed significant
$\pp$ and $\pl$ threshold enhancements in the radiative decays
$J/\psi \rightarrow \gamma \pp$~\cite{besppbar} and $J/\psi
\rightarrow \pl K^- + {\it c.c.}$~\cite{besplambda}, as well as in
$\psi(2S) \rightarrow \pl K^- + {\it c.c.}$~\cite{besplambda}.
More recently, a state $X(1835)$, decaying to $\pi^+\pi^-
\eta^{\prime}$, has been reported in $J/\psi \rightarrow \gamma
X$~\cite{x1835}. The state $X(1835)$ is consistent with the
enhancement at $\pp$ threshold if $J^{PC} = 0^{-+}$ is assumed. It
could be a glueball with large $\pp$ content in its wave function,
or a $\pp$ ``baryonium" resonance~\cite{zhu, zhu1,zhu2,zhu3,zhu4}, or other
possibilities~\cite{others00,others01,others02,others03,others04,others05,
others31,others32,others33}.

Baryon-antibaryon enhancements near threshold also appear in $B
\rightarrow \pp K$, $\Bz \rightarrow \pl \pi^-$, and $\Bu
\rightarrow \pp \pi^+$ decays~\cite{bfactory}. The increased
number of baryon-antibaryon enhancements near threshold attracts
more attention to study the possible existence of this kind of
exotics beyond the planned hadron spectrum~\cite{jafe}, or to
think about ``why don't multi-nucleons merge into single bags and
how do they merge?"~\cite{frank}. Experimentally, it will be very
important to search for the direct evidence of the formation of
baryon-antibaryon bound states.

In this letter, we consider the following processes
\begin{eqnarray}
& \pp & \rightarrow \gamma X^0 (X^0 \rightarrow \pi^+\pi^- \eta^{\prime}),
\label{eq:reaction1}\\
& \np & \rightarrow \gamma X^- (X^- \rightarrow \pi^0\pi^- \eta^{\prime}) ,
\label{eq:reaction2}
\end{eqnarray}
where the cold $\pbar$ beam pours into the liquid hydrogen or
helium target, and $\gamma$ rays are released as binding energy
once possible $\pp$ ($\np$) bound state is formed, $X^0$ ($X^-$)
can be reconstructed from decay  $\pi^+\pi^- \eta^{\prime}$
($\pi^0\pi^- \eta^{\prime}$). The $\gamma$ ray is monoenergetic
when an antiproton at rest is bounded with a proton or a neutron
in the target. For the $\pp$ enhancement $X(1860)$ observed by
BES,  in the case of an $S$-wave fit, the peak mass below $2m_p$
at $M(X(1860)) = 1859^{+3}_{-10}(\mbox{stat})^{+5}_{-25}
(\mbox{sys})$ MeV/$c^2$~\cite{besppbar} is given,  the
corresponding energy of $\gamma$ ray in
reaction~(\ref{eq:reaction1}) is $E_{\gamma} = 17.5$ MeV. While,
for the $X(1835)$ observed by BES in $J/\psi \rightarrow \gamma
\pi^+\pi^- \eta^{\prime}$ mode~\cite{x1835}, the energy of the
$\gamma$ ray is $E_{\gamma} = 42.4$ MeV if $X(1835)$ is also
formed by $\pp$ binding process. Certainly, $X(1860)$ and
$X(1835)$ may be the same resonance observed in different decay
channels~\cite{x1835,zhu,zhu1,zhu2,zhu3,zhu4}.  If this is true, only one peak in the
$\gamma$ spectrum is expected to be observed in
reaction~(\ref{eq:reaction1}), with $E_{\gamma}$ ranging from 10
to 100 MeV. This is the direct experimental test whether $X(1860)$
and $X(1835)$ are the same resonance formed by $\pp$ system.

There will be two kinds of dominant backgrounds in the above
reactions. The first one is the non-resonant $\pi^+\pi^-
\eta^{\prime}$ final states, the second one is the known regular
resonances below $\pp$ threshold, which are also allowed to be
produced at the same time in $\pp$ annihilation. For the first,
the energy distribution of the radiative $\gamma$ ray is different
from the monoenergetic $\gamma$ ray released as binding energy to
form a $\pp$ bound state.  However, the second background is more
dangerous since the radiative $\gamma$ ray is also monoenergetic,
which may dilute the search for the narrow $\pp$ bound states.
From this point of view, other techniques have to be employed to
distinguish signal from backgrounds, such as, a partial wave
analysis.

It is very interesting to look for $\pi^+\pi^- \eta^{\prime}$ in
$\pp$ annihilation at rest, in order to confirm the new
observation of $X(1835)$ state from the BES Collaboration. We find
that there were a few experiments at CERN Proton Synchrotron (PS)
aimed to search for the narrow state near $\pp$ threshold in
1970s~\cite{pscern}, but the published $\gamma$ ray spectrum are
only above 100 MeV with low statistics and low significance
(around 3$\sigma$ or below)~\cite{psresults}, a detailed review on
the nucleon-antinucleon scattering experiments performed at LEAR
of CERN can be found in reference~\cite{reportppbar}. The more
recent experiment is the Crystal Barrel detector at
LEAR/CERN~\cite{PS185} with good resolutions for both $\gamma$ ray
and charged track.  One can use the data from stopped $\pbar$
annihilation~\cite{nnbarCB} at this experiment to search for the
possible narrow bound states below the $\pp$ threshold.

Details of the Crystal Barrel detector are given in
reference~\cite{PS185}, here we give a short outline of its main
components for an easy reference. There are a barrel of 1380 CsI
crystals measuring energies and directions of $\gamma$-rays with
near 4$\pi$ solid angle coverage (0.95$\times$ 4$\pi$). It
provides high-efficiency photon detection with good energy and
spatial resolutions over an energy range from 20 MeV to 2000 MeV.
The typical energy resolution for photon with energy $E$ (in GeV)
is $\sigma_{E}/E = 2.5\%/E^{1/4}$, and $\sigma_{\theta,\phi} =
1.2^{\circ}$ in both polar and azimuthal angles. The mass
resolution is $\sigma = 10$~MeV/$c^2$ for $\pi^0 \to \gamma\gamma$
and 17~MeV for $\eta \to \gamma \gamma$. The energy of the
radiative $\gamma$ ray in reactions~(\ref{eq:reaction1})
and~(\ref{eq:reaction2}) ranges from 17 MeV to 100 MeV, the
corresponding relative resolutions are from 7\% to 4\%, which are
still good enough for detecting the proposed reactions. There are
two proportional chambers and a JET drift chamber with 23 layers
in which charged particles are identified and their curvatures are
measured in a 1.5 T homogeneous magnetic field. This gives a
momentum resolution of $\sigma_p / p = 2\%$ at 200 MeV/c, rising
to 7\% at 1 GeV/c, for charged pions.

We also note that a next-generation low-energy antiproton
facility, the Facility for Low-energy Antiproton and Ion Research
(FLAIR), is under discussion~\cite{flair} at GSI. It will provide
cooled beams at higher intensities and a factor of 10 or more
lower energy than the Antiproton Decelerator (AD) at CERN, the
effective $\pbar$ beam intensity is as large as $10^{10}$
$\pbar$/s~\cite{flair}.  FLAIR can extract cooled beam at energies
between 20 keV and 30 MeV, and cooled particles at rest or at
ultra-low eV energies. The experiment using antiprotons at such
facility supplies an unique opportunity for a direct and sensitive
investigation of the QCD structure of the $\pp$ or $\np$ system.
The detailed description of the development of these experimental
techniques can be found in reference~\cite{progress,progress1}.

In addition to the above proposed measurements, one can also
measure $\sigma(\pp \rightarrow \gamma X)$ at fine steps of
$\pbar$ beam energies very close to the threshold, these
measurements are helpful for studying of the production dynamics
of possible existing $\pp$ or $\np$ bound states.

In conclusion, we discussed the possibility of searching for
direct evidence of the existence of the $\pp$ bound states.  In
addition to the neutral candidate in $\pp$ channel, a charged
resonance $X^-$ is also proposed to be searched for in $\np$
channel. The Crystal Barrel data from  $\pp$ annihilation at rest
can be used to look for the inclusive $\gamma$ ray ($ \pp
\rightarrow \gamma X^0, X^0 \rightarrow \mbox{anything} $ ) or
exclusive channel ($ \pp  \rightarrow \gamma X^0, X^0 \rightarrow
\pi^+\pi^- \eta^{\prime}$) with monoenergy if the width of the
$\pp$ bound state is small. New techniques at the next-generation
$\pbar$ beams at FLAIR can be employed to study the structure of
the $X(1860)$ and $X(1835)$ in the future.

One of the authors (H. Li) would like to thank
Ulrich Wiedner and Bingsong Zou for helpful communications.
This work is supported in part by the National Natural Science Foundation
of China under contracts Nos. 10205017, 10575108, 10491303 and the Knowledge Innovation Project of
CAS under contract No. U-612 (IHEP).




\end{document}